\errorstopmode
\input amssym.def
\input amssym.tex

% Page layout

\magnification=\magstephalf
\hsize=14.0 true cm
\vsize=19 true cm
\hoffset=1.0 true cm
\voffset=2.0 true cm

\abovedisplayskip=12pt plus 3pt minus 3pt
\belowdisplayskip=12pt plus 3pt minus 3pt
\parindent=1.0em

% Fonts

\font\sixrm=cmr6
\font\eightrm=cmr8
\font\ninerm=cmr9

\font\sixi=cmmi6
\font\eighti=cmmi8
\font\ninei=cmmi9

\font\sixsy=cmsy6
\font\eightsy=cmsy8
\font\ninesy=cmsy9

\font\sixbf=cmbx6
\font\eightbf=cmbx8
\font\ninebf=cmbx9

\font\eightit=cmti8
\font\nineit=cmti9

\font\eightsl=cmsl8
\font\ninesl=cmsl9

\font\sixss=cmss8 at 8 true pt
\font\sevenss=cmss9 at 9 true pt
\font\eightss=cmss8
\font\niness=cmss9
\font\tenss=cmss10

\font\sixmib=cmmib6
\font\sevenmib=cmmib7
\font\eightmib=cmmib8
\font\ninemib=cmmib9
\font\tenmib=cmmib10

 at 12 true pt
 at 12 true pt
\font\bigrm=cmr10 at 12 true pt
 at 12 true pt
 at 12 true pt

 at 16 true pt
%\font\Bigsy=cmsy12 at 16 true pt
%\font\Bigex=cmex12 at 16 true pt
 at 16 true pt
\font\Bigrm=cmr12 at 16 true pt
 at 16 true pt
 at 16 true pt

\catcode`@=11
\newfam\ssfam
\newfam\mibfam

\def\tenpoint{\def\rm{\fam0\tenrm}%
    \textfont0=\tenrm \scriptfont0=\sevenrm \scriptscriptfont0=\fiverm
    \textfont1=\teni  \scriptfont1=\seveni  \scriptscriptfont1=\fivei
    \textfont2=\tensy \scriptfont2=\sevensy \scriptscriptfont2=\fivesy
    \textfont3=\tenex \scriptfont3=\tenex   \scriptscriptfont3=\tenex
    \textfont\itfam=\tenit                  \def\it{\fam\itfam\tenit}%
    \textfont\slfam=\tensl                  \def\sl{\fam\slfam\tensl}%
    \textfont\bffam=\tenbf \scriptfont\bffam=\sevenbf
                           \scriptscriptfont\bffam=\fivebf
                           \def\bf{\fam\bffam\tenbf}%
    \textfont\ssfam=\tenss \scriptfont\ssfam=\sevenss
                           \scriptscriptfont\ssfam=\sevenss
                           \def\ss{\fam\ssfam\tenss}%
    \textfont\mibfam=\tenmib \scriptfont\mibfam=\sevenmib
                             \scriptscriptfont\mibfam=\sevenmib
                             \def\mib{\fam\mibfam\tenmib}%
    \normalbaselineskip=13pt
    \setbox\strutbox=\hbox{\vrule height8.5pt depth3.5pt width0pt}%
    \let\big=\tenbig
    \normalbaselines\rm}

\def\ninepoint{\def\rm{\fam0\ninerm}%
    \textfont0=\ninerm      \scriptfont0=\sixrm
                            \scriptscriptfont0=\fiverm
    \textfont1=\ninei       \scriptfont1=\sixi
                            \scriptscriptfont1=\fivei
    \textfont2=\ninesy      \scriptfont2=\sixsy
                            \scriptscriptfont2=\fivesy
    \textfont3=\tenex       \scriptfont3=\tenex
                            \scriptscriptfont3=\tenex
    \textfont\itfam=\nineit \def\it{\fam\itfam\nineit}%
    \textfont\slfam=\ninesl \def\sl{\fam\slfam\ninesl}%
    \textfont\bffam=\ninebf \scriptfont\bffam=\sixbf
                            \scriptscriptfont\bffam=\fivebf
                            \def\bf{\fam\bffam\ninebf}%
    \textfont\ssfam=\niness \scriptfont\ssfam=\sixss
                            \scriptscriptfont\ssfam=\sixss
                            \def\ss{\fam\ssfam\niness}%
    \textfont\mibfam=\ninemib \scriptfont\mibfam=\sixmib
                            \scriptscriptfont\mibfam=\sixmib
                            \def\mib{\fam\mibfam\ninemib}%
    \normalbaselineskip=12pt
    \setbox\strutbox=\hbox{\vrule height8.0pt depth3.0pt width0pt}%
    \let\big=\ninebig
    \normalbaselines\rm}

\def\eightpoint{\def\rm{\fam0\eightrm}%
    \textfont0=\eightrm      \scriptfont0=\sixrm
                             \scriptscriptfont0=\fiverm
    \textfont1=\eighti       \scriptfont1=\sixi
                             \scriptscriptfont1=\fivei
    \textfont2=\eightsy      \scriptfont2=\sixsy
                             \scriptscriptfont2=\fivesy
    \textfont3=\tenex        \scriptfont3=\tenex
                             \scriptscriptfont3=\tenex
    \textfont\itfam=\eightit \def\it{\fam\itfam\eightit}%
    \textfont\slfam=\eightsl \def\sl{\fam\slfam\eightsl}%
    \textfont\bffam=\eightbf \scriptfont\bffam=\sixbf
                             \scriptscriptfont\bffam=\fivebf
                             \def\bf{\fam\bffam\eightbf}%
    \textfont\ssfam=\eightss \scriptfont\ssfam=\sixss
                             \scriptscriptfont\ssfam=\sixss
                             \def\ss{\fam\ssfam\eightss}%
    \textfont\mibfam=\eightmib \scriptfont\mibfam=\sixmib
                             \scriptscriptfont\mibfam=\sixmib
                             \def\mib{\fam\mibfam\eightmib}%
    \normalbaselineskip=10pt
    \setbox\strutbox=\hbox{\vrule height7.0pt depth2.0pt width0pt}%
    \let\big=\eightbig
    \normalbaselines\rm}

\def\tenbig#1{{\hbox{$\left#1\vbox to8.5pt{}\right.\n@space$}}}
\def\ninebig#1{{\hbox{$\textfont0=\tenrm\textfont2=\tensy
                       \left#1\vbox to7.25pt{}\right.\n@space$}}}
\def\eightbig#1{{\hbox{$\textfont0=\ninerm\textfont2=\ninesy
                       \left#1\vbox to6.5pt{}\right.\n@space$}}}

\font\sectionfont=cmbx10
\font\subsectionfont=cmti10

\def\figurecaptionfont{\ninepoint}
\def\tablecaptionfont{\ninepoint}
\def\footnotefont{\eightpoint}

% New count registers

\newcount\equationno
\newcount\bibitemno
\newcount\figureno
\newcount\tableno

\equationno=0
\bibitemno=0
\figureno=0
\tableno=0
%\advance\pageno by -1

% Footline

\footline={\ifnum\pageno=0{\hfil}\else
{\hss\rm\the\pageno\hss}\fi}

% Section macro

\def\section #1. #2 \par
{\vskip0pt plus .10\vsize\penalty-100 \vskip0pt plus-.10\vsize
\vskip 1.6 true cm plus 0.2 true cm minus 0.2 true cm
\global\def\equationlabel{#1}
\global\equationno=0
\leftline{\sectionfont #1. #2}\par
\immediate\write\terminal{Section #1. #2}
\vskip 0.7 true cm plus 0.1 true cm minus 0.1 true cm
\noindent}

% Subsection macro

\def\subsection #1 \par
{\vskip0pt plus 1.0 true cm\penalty-50 \vskip0pt plus-1.0 true cm
\vskip2.5ex plus 0.1ex minus 0.1ex
\leftline{\subsectionfont #1}\par
\immediate\write\terminal{Subsection #1}
\vskip1.0ex plus 0.1ex minus 0.1ex
\noindent}

% Appendix macro

\def\appendix #1. #2 \par
{\vskip0pt plus .10\vsize\penalty-100 \vskip0pt plus-.10\vsize
\vskip 1.6 true cm plus 0.2 true cm minus 0.2 true cm
\global\def\equationlabel{\hbox{\rm#1}}
\global\equationno=0
\leftline{\sectionfont Appendix #1. #2}\par
\immediate\write\terminal{Appendix #1. #2}
\vskip 0.7 true cm plus 0.1 true cm minus 0.1 true cm
\noindent}

%\def\appendix #1. #2 \par
%{\vskip0pt plus .20\vsize\penalty-100 \vskip0pt plus-.20\vsize
%\vskip 1.6 true cm plus 0.2 true cm minus 0.2 true cm
%\global\def\equationlabel{\hbox{\rm#1}}
%\global\equationno=0
%\leftline{\sectionfont Appendix #1. #2}\par
%\immediate\write\terminal{Appendix #1. #2}
%\vskip 0.7 true cm plus 0.1 true cm minus 0.1 true cm
%\noindent}

% Displayed equations

\def\equation#1{$$\displaylines{\qquad #1}$$}
\def\enum{\global\advance\equationno by 1
\hfill\llap{{\rm(\equationlabel.\the\equationno)}}}

\def\nexteq#1{\cr\noalign{\vskip#1}\qquad}

% Bibliography macro, references

\def\ifundefined#1{\expandafter\ifx\csname#1\endcsname\relax}

\def\ref#1{\ifundefined{#1}?\immediate\write\terminal{unknown reference
on page \the\pageno}\else\csname#1\endcsname\fi}

\newwrite\terminal
\newwrite\bibitemlist

\def\bibitem#1#2\par{\global\advance\bibitemno by 1
\immediate\write\bibitemlist{\string\def
\expandafter\string\csname#1\endcsname
{\the\bibitemno}}
\item{[\the\bibitemno]}#2\par}

\def\beginbibliography{
\vskip0pt plus .15\vsize\penalty-100 \vskip0pt plus-.15\vsize
\vskip 1.2 true cm plus 0.2 true cm minus 0.2 true cm
\leftline{\sectionfont References}\par
\immediate\write\terminal{References}
\immediate\openout\bibitemlist=biblist
\frenchspacing\parindent=1.8em
\vskip 0.5 true cm plus 0.1 true cm minus 0.1 true cm}

\def\endbibliography{
\immediate\closeout\bibitemlist
\nonfrenchspacing\parindent=1.0em}

% Figure and table captions

\def\figurecaption#1{\global\advance\figureno by 1
\narrower\figurecaptionfont Fig.~\the\figureno. #1}

\def\tablecaption#1{\global\advance\tableno by 1
\hfill{\tablecaptionfont Table~\the\tableno. #1}\hfill}

\def\thicktablerule{\hrule height0.8pt}
\def\thintablerule{\hrule height0.4pt}

\tenpoint

\immediate\openin\bibitemlist=biblist
\ifeof\bibitemlist\immediate\closein\bibitemlist
\else\immediate\closein\bibitemlist
\input biblist \fi

% current year and month

\def\thismonth{\ifcase\month\or
January\or February\or March\or April\or May\or June\or
July\or August\or September\or October\or November\or December\fi}

\input epsf
\epsfclipon

% Definitions and abbreviations

% Roman letters in math formulae

\def\rme{{\rm e}}
\def\rmO{{\rm O}}

% Real and integer numbers

\def\Re{{\rm Re}\,}

% Special relations and symbols

\def\proof{\noindent{\sl Proof:}\kern0.6em}

\def\frac#1#2{\hbox{$#1\over#2$}}
\def\dual{\mathstrut^*\kern-0.1em}

\def\lvec#1{\setbox0=\hbox{$#1$}
    \setbox1=\hbox{$\scriptstyle\leftarrow$}
    #1\kern-\wd0\smash{
    \raise\ht0\hbox{$\raise1pt\hbox{$\scriptstyle\leftarrow$}$}}
    \kern-\wd1\kern\wd0}
\def\rvec#1{\setbox0=\hbox{$#1$}
    \setbox1=\hbox{$\scriptstyle\rightarrow$}
    #1\kern-\wd0\smash{
    \raise\ht0\hbox{$\raise1pt\hbox{$\scriptstyle\rightarrow$}$}}
    \kern-\wd1\kern\wd0}
\def\slash#1{\setbox0=\hbox{$#1$}\setbox1=\hbox{$\kern1pt/$}
    #1\kern-\wd0\kern1pt/\kern-\wd1\kern\wd0}

% Lattice derivatives

\def\nabstar#1{{\nabla\kern0.5pt\smash{\raise 4.5pt\hbox{$\ast$}}
               \kern-5.5pt_{#1}}}

\def\drvstar#1{{\partial\kern0.5pt\smash{\raise 4.5pt\hbox{$\ast$}}
               \kern-6.0pt_{#1}}}

\def\ldrvstar#1{{\lvec{\,\partial}\kern-0.5pt\smash{\raise 4.5pt\hbox{$\ast$}}
               \kern-5.0pt_{#1}}}

% Units

\def\MSbar{\overline{\rm MS\kern-0.5pt}\kern0.5pt}

% Constants

% Fields

\def\psibar{\overline{\psi}{\vphantom{\psi}}\kern-0.6pt}
\def\chibar{\overline{\chi}{\vphantom{\chi}}\kern-0.6pt}

% Dirac matrices

\def\dirac#1{\gamma_{#1}}
\def\diracstar#1#2{
    \setbox0=\hbox{$\gamma$}\setbox1=\hbox{$\gamma_{#1}$}
    \gamma_{#1}\kern-\wd1\kern\wd0
    \smash{\raise4.5pt\hbox{$\scriptstyle#2$}}}

% Gauge group

\def\SUthree{{\rm SU(3)}}

\def\tr{{\rm tr}}

% Action

\def\SG{S_{\rm G}}
\def\Sw{S_{\rm w}}
\def\Spf#1{S_{{\rm pf},#1}}
\def\Spft#1{\tilde{S}_{{\rm pf},#1}}

\def\lp{{\cal C}}
\def\lps#1{{\cal S}_{#1}}
\def\csw{c_{\rm sw}}
\def\cG{c_{\hbox{\sixrm G}}}
\def\cF{c_{\hbox{\sixrm F}}}

% Dirac operator

\def\Dm{D}
\def\Dmdag{\Dm^{\dagger}\kern-1pt}
\def\Dms{D_s}
\def\Dmsdag{D_s\kern-2pt\vphantom{D}^{\dagger}\kern-1pt}

% Even-odd preconditioning

\def\Dee{D_{\rm ee}}
\def\Doo{D_{\rm oo}}
\def\Deo{D_{\rm eo}}
\def\Doe{D_{\rm oe}}
\def\Dmh{\hat{D}}
\def\Dmhdag{\Dmh^{\dagger}\kern-1pt}
\def\Dmsh{\hat{D}_s}
\def\Dmshdag{\hat{D}_s\kern-2pt\vphantom{D}^{\dagger}\kern-1pt}

% HMC

\def\Pacc{P_{\rm acc}}

\def\Ws{W_s}
\def\Ntr{N_{\rm tr}}

% Masses

\def\mpi{m_{\pi}}
\def\mK{m_K}

% Misc

\def\eps{\epsilon}
\def\Ebar{\kern1.5pt\overline{\kern-1.5ptE\kern-0.5pt}\kern0.5pt}
\def\cvec#1{\kern-0.5pt\vec{\kern0.5pt #1}}

%
%\vbox{\vskip0.0cm}
\rightline{CERN-PH-TH/2012-161}
\vskip1.2cm 
\centerline{\Bigrm
Lattice QCD with open boundary conditions}

\vskip0.3 true cm
\centerline{\Bigrm and twisted-mass reweighting}
\vskip 0.6 true cm
\centerline{\bigrm Martin L\"uscher and Stefan Schaefer}
\vskip1.5ex
\centerline{{\it CERN, Physics Department, 1211 Geneva 23, Switzerland}}
\vskip 0.8 true cm
\thintablerule
\vskip 2.0ex
\ninepoint
\leftline{\bf Abstract}
\vskip 1.0ex\noindent
Lattice QCD simulations at small lattice spacings and 
quark masses close to their physical values are 
technically challenging.
In particular, the simulations can get trapped in the 
topological charge sectors of field space or
may run into instabilities triggered
by accidental near-zero modes of the lattice Dirac operator.
As already noted in ref.~[\ref{openQCD}], the first problem is bypassed 
if open boundary conditions are imposed in the time direction, while the
second can potentially be overcome through twisted-mass
determinant reweighting [\ref{TMRW}]. 
In this paper, we show that twisted-mass reweighting works out
as expected in QCD with open boundary conditions
and 2+1 flavours of O($a$) improved Wilson quarks.
Further algorithmic improvements are tested as well 
and a few physical quantities are computed for illustration.
\vskip 2.0ex
\thintablerule

\tenpoint

\vskip-0.3cm

\section 1. Introduction

To be able to control the systematic errors in lattice QCD computations,
simulations of lattices with spacing smaller than $0.05$ fm and 
spatial extent of at least $4$ fm have to be performed. Moreover,
the quark masses should ideally be set to 
their physical values in these simulations. 

An obstacle to progress along these lines
is the well-established fact that all known simulation
algorithms tend to get trapped in the sectors of gauge fields 
with fixed topological charge [\ref{DelDebbioTauQ}--\ref{Villasimius}].
So far no remedy against
this loss of ergodicity was found, but the problem can be
bypassed by choosing open boundary conditions for the gauge
field in the time direction [\ref{openQCD}]. 
The topological charge can then flow in and out 
of the lattice through its boundaries, while the physical states
and the Hamiltonian are unchanged.

In this paper, the formulation of lattice QCD proposed by Wilson 
[\ref{Wilson}] is adopted, with counterterms added to cancel
the leading effects in the lattice spacing $a$ [\ref{SW},\ref{SFimp}].
This version of the lattice theory has many desirable
properties and is relatively easy to simulate. Chiral symmetry
is however violated by effects of order $a^2$ and the spectrum of the
Dirac operator is therefore not protected against accidental near-zero
modes. Such modes can give rise to instabilities
in simulations based on the HMC algorithm [\ref{HMC}], which may, 
in the worst case, compromise the correctness of the simulations. 

At fixed gauge coupling and quark masses, 
near-zero modes tend to be less frequent the
larger the lattice volume $V$ is, because
the width of the distribution of the lowest eigenvalue of the Dirac operator 
decreases approximately like $V^{-1/2}$ [\ref{Stability}--\ref{RMTIV}].
While the shrinking of the width of the eigenvalue distribution has
a stabilizing effect on large lattices, the twisted-mass 
determinant reweighting proposed in ref.~[\ref{TMRW}] avoids
the problem from the beginning through an intermediate infrared 
regularization of the quark determinant.

Encouraging first tests of this method 
were recently reported by Miao et al.~[\ref{MiaoEtAl}].
Here we shall present the
results of a more complete study that includes simulations
of QCD with 2+1 flavours of quarks 
at a point in parameter space previously considered by the 
PACS-CS collaboration [\ref{AokiEtAlI},\ref{AokiEtAlII}],
where the quark masses are practically equal to their 
physical values. The simulations we have performed for
these tests are also the first ones of full QCD with
open boundary conditions. Moreover, 
an effort was made to improve the efficiency and robustness
of the simulation algorithm by combining various known
techniques (see sections 3 and 4).
All simulations reported in this paper were performed using the
publicly available {\tt openQCD} program package [\ref{OQCD}].

\section 2. Twisted-mass determinant reweighting

\vskip-2.5ex

\subsection 2.1 Lattice theory

As already mentioned, we consider lattice QCD with 
O($a$)-improved Wilson quarks. The up and down
quarks are assumed to be mass-degenerate and
are referred to as the light quarks. 
There could be any number of heavier quarks, but 
the strange quark is the only one included in 
the simulations reported later.

The basic setup of the lattice theory and the notation employed
are as in ref.~[\ref{openQCD}]. In particular, open boundary
conditions are imposed in the time direction and the
(four-dimensional hypercubic) lattice is assumed to have 
time-like extent $T$ and spatial size $L$. For notational convenience,
the lattice spacing is set to unity.
As for the gauge
action, we slightly depart from ref.~[\ref{openQCD}] 
and replace the Wilson action by a more general expression,
which includes the tree-level Symanzik-improved and the Iwasaki
action (see appendix A).

\subsection 2.2 Determinant regularization

Let $\Dm$ be the up quark lattice Dirac operator as it appears
in the lattice action.
The operator thus includes the (ordinary) mass term and the required
O($a$) counterterms.
In ref.~[\ref{TMRW}], two kinds of twisted-mass regularizations
of the light quark determinant were proposed, which amount
to replacing
\equation{
  \det\{\Dmdag\Dm\}\to\det\{\Dmdag\Dm+\mu^2\}
  \enum
}
or
\equation{
  \det\{\Dmdag\Dm\}\to\det\{(\Dmdag\Dm+\mu^2)^2(\Dmdag\Dm+2\mu^2)^{-1}\}
  \enum
}
respectively. The twisted mass parameter $\mu>0$ provides
the desired infrared regularization and is usually set to
a value on the order of the light quark mass.

With the regularization in place, the ensembles of 
gauge-field configurations generated in the simulations must be
reweighted in order to obtain the correct expectation 
values of the observables of interest.
The reweighting factors in the two cases are
\equation{
  W_1=\det\{\Dmdag\Dm(\Dmdag\Dm+\mu^2)^{-1}\},
  \enum
  \nexteq{2.5ex}
  W_2=\det\{\Dmdag\Dm(\Dmdag\Dm+2\mu^2)(\Dmdag\Dm+\mu^2)^{-2}\}.
  \enum
}
Both factors are ratios of quark determinants, which 
can be estimated stochastically with a modest computational effort
(see subsect.~2.4).

Determinant reweighting usually becomes inefficient on large lattices,
but as explained in ref.~[\ref{TMRW}], the reweighting factors $W_1$ 
and $W_2$ are not expected to fluctuate wildly if $\mu$ is chosen
appropriately. 
The second form, eq.~(2.2), of the determinant regularization potentially
fares better in this respect, because
the contribution of the (very many) high modes of the Dirac operator to 
the reweighting factor is more strongly suppressed
than in the case of the first form.

In our empirical studies, we found that $W_2$ in fact tends to 
fluctuate less than $W_1$, although the behaviour of the two
factors is not qualitatively different
and moreover depends on the value of $\mu$. 
For simplicity, and since this is the method
used in the test runs reported later,
we shall from now on focus on the regularization
(2.2) of the quark determinant.

\subsection 2.3 Even-odd preconditioned version

Twisted-mass determinant reweighting easily combines
with the even-odd preconditioning of the Dirac operator.
Let 
\equation{
  \Dm=\pmatrix{\Dee & \Deo \cr
               \noalign{\vskip1ex}
               \Doe & \Doo \cr}
  \enum
}
be the block decomposition of the Dirac operator with respect to
an ordering of the lattice points $x$, where the even ones
(those with even $x_0+x_1+x_2+x_3$) come first. In practice
the blocks on the diagonal are always invertible so that
the even-odd preconditioned operator,
\equation{
  \Dmh=\Dee-\Deo(\Doo)^{-1}\Doe,
  \enum
}
is well defined. Note that $\Dmh$ acts on quark fields defined
on the even lattice points.

When even-odd preconditioning is used, eqs.~(2.2) and (2.4) get replaced by
\equation{
 \det\{\Dmdag\Dm\}\to\det\{(\Doo)^2\}
 \det\{(\Dmhdag\Dmh+\mu^2)^2(\Dmhdag\Dmh+2\mu^2)^{-1}\},
 \enum
 \nexteq{2.5ex}
 \hat{W}_2=\det\{\Dmhdag\Dmh(\Dmhdag\Dmh+2\mu^2)(\Dmhdag\Dmh+\mu^2)^{-2}\}.
 \enum
}
Note that the twisted mass term is added 
on the even sites of the lattice only.
The regularizations (2.2) and (2.7) are therefore not the same.

\subsection 2.4 Computation of the reweighting factor

An unbiased stochastic estimator for the reweighting factor $W_2$
is given by
\equation{
  W_{2,N}={1\over N}\sum_{k=1}^N
  \exp\{-\mu^4(\eta_k,(\Dmdag\Dm)^{-1}(\Dmdag\Dm+2\mu^2)^{-1}\eta_k)\},
  \enum
}
where $\eta_1,\ldots,\eta_N$ are random quark fields
with normal distribution and the bracket $(\cdot\,,\cdot)$ denotes the 
obvious scalar product of such fields.
The reweighting factor $\hat{W}_2$ 
can be similarly estimated by replacing $\Dm$ by $\Dmh$
and by restricting the random fields to the even sites of the lattice.

In practice, the number $N$ of random fields is 
usually taken to be in the range
from, say, $12$ to $48$.
The deviation $|W_{2,N}-W_2|$ is then often smaller than
the statistical fluctuations of $W_2$,
in which case the stochastic estimation of the reweighting factor 
does not lead to enhanced statistical errors.
More accurate determinations of the reweighting factor
(along the lines of ref.~[\ref{MRW}], for example) 
may however be required if observables sensitive to the low modes
of the Dirac operator are considered and if there is
an appreciable probability for the operator to 
have exceptionally small eigenvalues.

\section 3. Frequency splitting of the quark determinant

The numerical integration of the molecular-dynamics equations
can be a source of instability in the HMC algorithm
even at relatively large quark masses. Empirically one knows that a
frequency splitting of the quark determinant 
through mass shifts [\ref{Hasenbusch}--\ref{UrbachEtAl}] 
or a domain decomposition of the Dirac operator [\ref{DDHMC}] 
has a stabilizing effect and thus allows the equations 
to be integrated with larger step sizes than would otherwise be
possible. 

In this section, we describe a particular 
frequency-splitting scheme that 
naturally goes together with the twisted-mass determinant reweighting. 
The scheme has further merits and 
performed very well in all simulations reported later.
For simplicity, we only discuss the case without
even-odd preconditioning, but all results extend
to the preconditioned quark determinants with the obvious modifications.

\subsection 3.1 Factorization of the light-quark determinant

Let $\mu_0,\ldots,\mu_n$ be a set of twisted mass parameters satisfying
\equation{
  \mu_0=\mu,\quad\mu_0<\mu_1<\ldots<\mu_n,
  \enum
}
where $\mu$ is the regulator mass used for the determinant reweighting.
The regularized light-quark determinant on the right of eq.~(2.2) may 
then be written in the factorized form 
[\ref{Hasenbusch},\ref{HasenbuschJansen}]
\equation{
  \det\left\{\Dmdag\Dm+\mu_n^2\right\}
  \det\left\{{\Dmdag\Dm+\mu_0^2\over\Dmdag\Dm+2\mu_0^2}\right\}
  \prod_{k=0}^{n-1}
  \det\left\{{\Dmdag\Dm+\mu_k^2\over\Dmdag\Dm+\mu_{k+1}^2}\right\}.
  \enum
}
Each factor in this product of determinants may be represented through
a pseudo-fermion functional integral. In total 
$n+2$ independent pseudo-fermion fields,
$\tilde{\phi}_0$ and $\phi_0,\phi_1,\ldots,\phi_n$, 
with actions
\equation{
  \Spft{0}=\bigl(\tilde{\phi}_0,(\Dmdag\Dm+2\mu_0^2)(\Dmdag\Dm+\mu_0^2)^{-1}
                 \tilde{\phi}_0\bigr),
  \enum
  \nexteq{2.5ex}
  \Spf{k}=\bigl(\phi_k,(\Dmdag\Dm+\mu_{k+1}^2)
                       (\Dmdag\Dm+\mu_k^2)^{-1}\phi_k\bigr),
  \quad k=0,\ldots,n-1,
  \enum
  \nexteq{2.5ex}
  \Spf{n}=\bigl(\phi_n,(\Dmdag\Dm+\mu_n^2)^{-1}\phi_n\bigr),
  \enum
}
need to be introduced for this representation.

When written as products over the eigenvalues of $\Dmdag\Dm$,
the determinants in eq.~(3.2) are seen to be dominantly dependent
on the eigenvalues in spectral intervals that are roughly
delimited by the twisted masses $\mu_0,\ldots,\mu_n$.
To some extent, at least,
the factorization thus achieves a frequency splitting of 
the light-quark determinant.
While such factorizations are known to
stabilize the HMC algorithm,
there is currently no solid theoretical understanding
of why this is so. As a consequence, the choice of 
the twisted masses is, in practice, 
poorly guided and may require some fine-tuning [\ref{UrbachEtAl}].

In the course of our algorithmic studies, we found that
frequency splittings where $\mu_n\simeq1$ and
\equation{
  \mu_k\simeq0.1\times\mu_{k+1},\quad k=0,\ldots,n-1,
  \enum
}
gave good results in all cases considered.
Moreover, it is our experience that a fine-tuning of the 
masses is then not required. The rule (3.6) amounts to 
splitting the spectral range of the Dirac operator 
in equal segments on a log scale
and is therefore referred to as ``log-scale frequency splitting''.
Note that the rule implicitly fixes the number of twisted
masses as a function of the reweighting mass $\mu=\mu_0$.

\subsection 3.2 Strange quark determinant

While the physical strange quark is much heavier than the light quarks,
the condition number of the strange-quark lattice Dirac operator $\Dms$ is 
not small in practice and
a frequency splitting of strange-quark determinant thus
seems advisable. Such splittings are naturally obtained when
the RHMC algorithm [\ref{RHMCI},\ref{RHMCII}] is employed for the 
strange quark.

The version of the RHMC algorithm used here 
essentially coincides with the one described in sect.~2.6 of 
ref.~[\ref{LesHouches}]. The starting point is the 
factorization
\equation{
  \det\Dms=\Ws\det R^{-1}
  \enum
}
of the strange-quark determinant, where
\equation{
  R=C\prod_{k=0}^{m-1}{\Dmsdag\Dms+\omega_k^2\over\Dmsdag\Dms+\nu_k^2}
  \enum
}
denotes the Zolotarev optimal rational approximation 
[\ref{Achiezer}] of degree $m$ of the operator 
$(\Dmsdag\Dms)^{-1/2}$ and
\equation{
  \Ws=\det(\Dms R)
  \enum
}
the reweighting factor needed to correct for the approximation error.
For a specified degree $m$ and spectral range 
of  $\Dmsdag\Dms$ in which the approximation $R$ is to have
the least possible error, 
the constant $C$ and the twisted masses 
\equation{
  \nu_0<\omega_0<\nu_1<\ldots<\omega_{m-1}
  \enum
}
are uniquely determined. The latter typically range over a few orders
of magnitude and are about equally spaced on a log scale.

A further factorization of the strange-quark determinant 
is now obtained by breaking up
the Zolotarev rational function (3.8) into 
two or more factors of the form
\equation{
  R_{l,j}=\prod_{k=l}^{j}
  {\Dmsdag\Dms+\omega_k^2\over\Dmsdag\Dms+\nu_k^2}.
  \enum
}
If $m=12$, for example, a possible factorization is
\equation{
  \det R^{-1}=\hbox{constant}\times\det\{R_{0,4}^{-1}\}
  \det\{R_{5,8}^{-1}\}\det\{R_{9,11}^{-1}\}.
  \enum
}
Each factor $\det\{R_{l,j}^{-1}\}$
is then represented through an integral over a pseudo-fermion
field $\phi_{l,j}$ with action
\equation{
  \Spf{l,j}=(\phi_{l,j},R_{l,j}\phi_{l,j}).
  \enum
}
In view of the strong ordering of the twisted masses $\omega_k$
and $\nu_k$,
a frequency splitting of the strange-quark determinant is achieved 
in this way,
very much akin to the one of the light-quark determinant discussed
in subsect.~3.1.

\subsection 3.3 Molecular-dynamics forces

The molecular-dynamics force fields
that derive from the
actions $\Spf{k}$ are given by
\equation{
  F_k(x,\mu)^a=-2(\mu_{k+1}^2-\mu_k^2)
  \kern1pt\Re(\chi_k,\dirac{5}\partial^a_{x,\mu}D\psi_k),
  \quad k=0,\ldots,n-1,
  \enum
  \nexteq{2.5ex}
  F_n(x,\mu)^a=-2\kern1pt\Re(\chi_n,\dirac{5}\partial^a_{x,\mu}D\psi_n),
  \enum
}
where $\partial^a_{x,\mu}$ denotes the partial
derivative with respect to the link variable $U(x,\mu)$ 
in direction of the $\SUthree$ generator $T^a$ and
\equation{
  \psi_k=(D+i\mu_k\dirac{5})^{-1}\dirac{5}\phi_k,
  \qquad
  \chi_k=(D-i\mu_k\dirac{5})^{-1}\dirac{5}\psi_k.
  \enum
}
On physically large lattices, 
the force fields tend to be strongly ordered
in magnitude. In particular, 
in the case of log-scale frequency splitting,
the force $F_k$ is about $10$ times smaller than $F_{k+1}$
and the force $\tilde{F}_0$
deriving from the action $\Spft{0}$ is smaller
than $F_0$ by roughly another order of magnitude. 
When integrating the molecular-dynamics equations, the pseudo-fermion
forces can thus be integrated with different integration step sizes
without compromising the accuracy of the integration
[\ref{SextonWeingarten}].

Essentially the same comments apply to the forces deriving from the 
strange-quark pseudo-fermion actions (3.13). When the rational
functions are expanded in partial fractions, the contributions
of the fractions to the force are actually again given by eqs.~(3.14),(3.16)
except for a change in the proportionality factor and
the fact that the Dirac operator $\Dm$ is replaced by
$\Dms$.

\section 4. Integration of the molecular-dynamics equations

The numerical integration of the molecular-dynamics equations
can be accelerated by using the improved integrators proposed by
Omelyan, Mryglod and Folk [\ref{Omelyan}] and a 
locally deflated solver for the lattice Dirac equation
[\ref{DFLI},\ref{DFLII}]. In the following subsections, we
briefly describe the implementation of these improvements
in our simulations.

\subsection 4.1 Evolution equations

The molecular-dynamics equations
\equation{
  \partial_t\pi(x,\mu)=-T^a\partial^a_{x,\mu}S(U),
  \enum
  \nexteq{2.0ex}
  \partial_tU(x,\mu)=\pi(x,\mu)U(x,\mu),
  \enum
}
evolve the gauge field $U(x,\mu)$ and its momentum 
$\pi(x,\mu)=\pi(x,\mu)^aT^a$
as a function of the molecular-dynamics time $t$. 
If integrated exactly, the evolution preserves the 
Hamilton function
\equation{
  H(\pi,U)=\frac{1}{2}(\pi,\pi)+S(U),
  \qquad
  (\pi,\pi)=\sum_{x,\mu}\pi(x,\mu)^a\pi(x,\mu)^a.
  \enum
}
In these formulae, $S(U)$ stands for the total action, i.e.~the sum of the 
gauge action,
the pseudo-fermion actions and the terms proportional to 
$\ln\{\Doo\}$ and $\ln\{(\Dms)_{\rm oo}\}$, which
need to be included if even-odd preconditioning
is used (cf.~subsect.~2.3; the dependence of the action on the
pseudo-fermion fields is suppressed for simplicity).

\subsection 4.2 Elementary integrators

The numerical integration schemes used in our simulations
are based on the leapfrog integrator, 
the 2nd order Omelyan-Mryglod-Folk (OMF) integrator and a particular 
4th order OMF scheme.
All these integrators are sequences of the elementary update steps
\equation{
  {\cal I}_{\pi}(\eps):\;\pi\to\pi-\eps F,
  \enum
  \nexteq{2.5ex}
  {\cal I}_{U}(\eps):\;U\to\rme^{\eps\pi}U,
  \enum
}
where $\eps$ denotes the time step size and 
$F(x,\mu)=F(x,\mu)^aT^a$ the molecular-dynamics force integrated in the step.
The leapfrog integrator, for example, amounts to applying
the combination
\equation{
  {\rm LPFR}(\eps)
  ={\cal I}_{\pi}(\frac{1}{2}\eps)
   {\cal I}_{U}(\eps)
   {\cal I}_{\pi}(\frac{1}{2}\eps)
  \enum
} 
to the fields, while the 2nd order OMF integrator,
\equation{
  {\rm OMF}_2(\eps)
  ={\cal I}_{\pi}(\lambda\eps){\cal I}_{U}(\frac{1}{2}\eps)
   {\cal I}_{\pi}((1-2\lambda)\eps)
   {\cal I}_{U}(\frac{1}{2}\eps){\cal I}_{\pi}(\lambda\eps)
  \enum
} 
updates the gauge field in two steps and depends on a tunable parameter
$\lambda$. 
In the case of the 4th order OMF integrator, ${\rm OMF}_4(\eps)$, 
there are 5 update steps, with step sizes 
given by eqs.~(63) and (71) in ref.~[\ref{Omelyan}],
and no tunable parameters.

\subsection 4.3 Hierarchical integration

If the force $F$ is set to the one deriving from the total action, 
the $n$-fold application of one of the elementary integrators
integrates the molecular-dynamics equations from time 0 to some
later time $\tau=n\eps$. In practice, a hierarchical integration
scheme is used, where the different forces are integrated with
different time step sizes [\ref{SextonWeingarten}].

Hierarchical integrators are constructed recursively in a way that
is best explained by considering a simple case. Suppose the
total action $S=S_0+S_1$ is a sum of two terms that give rise
to the forces $F_0$ and $F_1$.
The construction then starts from
an integrator of the form
\equation{
  {\cal J}_1(\tau)=\bigl\{
  \left.{\rm OMF}_2(\eps_1)\right|_{F\to F_1}\bigr\}^{n_1},
  \qquad\eps_1=\tau/n_1,
  \enum
}
for the molecular-dynamics equations in which the total action
is replaced by $S_1$.
For a given integration time $\tau$, 
one has the choice of the elementary
integrator (${\rm OMF}_2$ in this case) at this level and 
of the number $n_1$ of times the latter is applied. 
Note that ${\cal J}_1(\tau)$ is just a sequence of 
update steps ${\cal I}_{\pi}(\eps)$ and ${\cal I}_{U}(\eps)$ with
varying step sizes $\eps$ proportional to $\eps_1$. 

The force $F_0$ may now be included in the molecular-dynamics evolution
by replacing all instances ${\cal I}_{U}(\eps)$
of the gauge-field update steps by an integrator like
\equation{
  {\cal J}_0(\eps)=\bigl\{
  \left.{\rm OMF}_4(\eps_0)\right|_{F\to F_0}\bigr\}^{n_0},
  \qquad\eps_0=\eps/n_0,
  \enum
}
which integrates $F_0$ from the current time $t$ to $t+\eps$.
At this level, one again has the choice of the elementary integrator
and the number of times it is applied.
The integrator obtained in this way integrates $F_0$ and $F_1$ with 
average step sizes equal to $\tau/(10n_0n_1)$ and $\tau/(2n_1)$,
respectively.

When the total action is a sum
of $n$ terms, a hierarchical integrator with $n$ levels is 
required if the associated forces are to be integrated with
different step sizes. The construction of the integrator always
starts from the top level, where the smallest forces are integrated, 
and proceeds to the lower levels recursively by replacing the gauge-field
update steps by a power of an elementary integrator.
For a complete description of the integration scheme, 
the integration time $\tau$,
the list of elementary integrators, the numbers $n_0,n_1,\ldots$
and the forces integrated at each
level must be specified.

\subsection 4.4 Deflation acceleration

The frequency splitting of the quark determinant
tends to give rise to a fairly large number of 
pseudo-fermion forces in the molecular-dynamics equations.
When some or all of these forces need to be computed,
the twisted-mass lattice Dirac equation must be solved 
several times. 

There is a range of algorithms
that allow the lattice Dirac equation to be solved efficiently.
In particular, the forces deriving from the rational-function actions
(3.13) can be computed using a multi-shift
conjugate-gradient (CG) algorithm [\ref{CGM}]. The 
CG algorithm is also suitable for the solution of the 
Dirac equation at large quark masses, but
at small and intermediate masses, the equation can be solved
much more rapidly using the GCR solver,
local deflation [\ref{DFLI}] and
a preconditioner based on the 
Schwarz alternating procedure (SAP) [\ref{SAP}].

Local deflation requires a deflation subspace 
to be generated before the integration of the molecular-dynamics
equations starts and to be kept up-to-date
in the course of the integration [\ref{DFLII}].
This overhead is however rapidly amortized
along the molecular-dynamics trajectories
if there are several pseudo-fermion forces, where
the use of the deflated solver is highly profitable
(as is the case at small quark masses).

\subsection 4.5 Remark on solver tolerances

Whichever iterative procedure is used for the solution
of the lattice Dirac equation, the algorithm is stopped
as soon as the residue of the calculated approximate
solution has decreased by some factor $\delta$. 
It is tempting to relax the tolerance $\delta$ as one
proceeds from the larger to the smaller forces,
since the latter are small corrections to 
the total force and therefore need not be computed as accurately
as the large forces [\ref{RHMCII}]. 

This argumentation however ignores the fact
that the deviations of the computed from the exact solutions
depend on the condition number of the lattice Dirac operator.
In the case of the fields (3.16), for example,
simple norm estimates actually suggest that the relative error
of the calculated fields $\psi_k$ and $\chi_k$ 
scale proportionally to $\delta/\mu_k$ and
$\delta/\mu_k^2$, respectively. Such rigorous estimates tend to be too
pessimistic, but they show that a loosening of the solver tolerances 
risks to compromise the accuracy (and thus
the stability) of the numerical integration of the 
molecular-dynamics equations.

\section 5. Algorithm stability and performance

One of the principal goals in this paper is to find out
whether twisted-mass determinant reweighting works out 
on large lattices and at quark masses close 
to their physical values. The simulations reported below
serve to study this question, but also provide a test
of the simulation algorithm described in sections $3$ and $4$.

\topinsert
%Blanke Zahl
\newdimen\digitwidth
\setbox0=\hbox{\rm 0}
\digitwidth=\wd0
\catcode`@=\active
\def@{\kern\digitwidth}
\tablecaption{Lattice parameters} 
\vskip-3.0ex
$$\vbox{\settabs\+&%                  
                  xxxxx&x&           Run
                  xxxxxx&xx&%         Lattice
                  xxxxxxx&xx&%        SG
                  xxxx&xx&%           beta
                  xxxxxxxx&xx&%       csw
                  xxxxxxxxxx&xx&%     kappa_u
                  xxxxxxxxx&\cr%     kappa_s
\thicktablerule
\vskip1.2ex
                \+& \hfill Run\hfill
                 && \hfill Lattice\hfill
                 && \hfill $\SG$\hfill
                 && \hfill $\beta$\hfill
                 && \hfill $\csw$\hfill
                 && \hfill $\kappa_u$\hfill
                 && \hfill $\kappa_s$\hfill
                 &\cr
\vskip1.0ex
\thintablerule
\vskip1.2ex
  \+& \hfill $D_6$\hfill
  &&  \hfill $48\times24^3$\hfill
  &&  \hfill Wilson\hfill
  &&  \hfill $5.3$\hfill
  &&  \hfill $1.90952$\hfill
  &&  \hfill $0.136350$\hfill
  &&  \hfill $-$\hfill
  &\cr
\vskip0.3ex
  \+& \hfill $E_8$\hfill
  &&  \hfill $64\times32^3$\hfill
  &&  \hfill Wilson\hfill
  &&  \hfill $5.3$\hfill
  &&  \hfill $1.90952$\hfill
  &&  \hfill $0.136417$\hfill
  &&  \hfill $-$\hfill
  &\cr
\vskip0.3ex
  \+& \hfill $I_1$\hfill
  &&  \hfill $64\times32^3$\hfill
  &&  \hfill Iwasaki\hfill
  &&  \hfill $1.9$\hfill
  &&  \hfill $1.71500$\hfill
  &&  \hfill $0.137740$\hfill
  &&  \hfill $0.136600$\hfill
  &\cr
\vskip0.3ex
  \+& \hfill $I_2$\hfill
  &&  \hfill $64\times32^3$\hfill
  &&  \hfill Iwasaki\hfill
  &&  \hfill $1.9$\hfill
  &&  \hfill $1.71500$\hfill
  &&  \hfill $0.137796$\hfill
  &&  \hfill $0.136634$\hfill
  &\cr
\vskip1.2ex
\thicktablerule
}
$$
\vskip-2.0ex
\endinsert

\topinsert
%Blanke Zahl
\newdimen\digitwidth
\setbox0=\hbox{\rm 0}
\digitwidth=\wd0
\catcode`@=\active
\def@{\kern\digitwidth}
\tablecaption{Lattice spacing and meson masses} 
\vskip-3.0ex
$$\vbox{\settabs\+&%
                  xxxxx&xx&%           Run
                  xxxxxxx&xx&%       a
                  xxxxxxx&xx&%       L
                  xxxxxxxxxx&xx&%    mpi
                  xxxxxxxxxx&xx&%    mK
                  xxxxxx&xx&%      mpi*L
                  xxxxxxxxxx&\cr%     Reference
\thicktablerule
\vskip1.2ex
                \+& \hfill Run\hfill
                 && \hfill $a$\kern2pt[fm]\hfill
                 && \hfill $L$\kern2pt[fm]\hfill
                 && \hfill $\mpi$\kern2pt[MeV]\hfill
                 && \hfill $\mK$\kern2pt[MeV]\hfill
                 && \hfill $\mpi L$\hfill
                 && \hfill Reference\hfill
                 &\cr
\vskip1.0ex
\thintablerule
\vskip1.2ex
  \+& \hfill $D_6$\hfill
  &&  \hfill $0.066$\hfill
  &&  \hfill $1.6$\hfill
  &&  \hfill $311$\hfill
  &&  \hfill $-$\hfill
  &&  \hfill $2.5$\hfill
  &&  \hfill [\ref{AlphaScale}]\hfill
  &\cr
\vskip0.3ex
  \+& \hfill $E_8$\hfill
  &&  \hfill $0.066$\hfill
  &&  \hfill $2.1$\hfill
  &&  \hfill $191$\hfill
  &&  \hfill $-$\hfill
  &&  \hfill $2.0$\hfill
  &&  \hfill [\ref{AlphaScale}]\hfill
  &\cr
\vskip0.3ex
  \+& \hfill $I_1$\hfill
  &&  \hfill $0.090$\hfill
  &&  \hfill $2.9$\hfill
  &&  \hfill $215$\hfill
  &&  \hfill $524$\hfill
  &&  \hfill $3.1$\hfill
  &&  \hfill [\ref{AokiEtAlI},\ref{AokiEtAlII}]\hfill
  &\cr
\vskip0.3ex
  \+& \hfill $I_2$\hfill
  &&  \hfill $0.090$\hfill
  &&  \hfill $2.9$\hfill
  &&  \hfill $135$\hfill
  &&  \hfill $498$\hfill
  &&  \hfill $2.0$\hfill
  &&  \hfill [\ref{AokiEtAlII}]\hfill
  &\cr
\vskip1.2ex
\thicktablerule
}
$$
\vskip-2.0ex
\endinsert

\subsection 5.1 Lattice parameters

The parameters of the lattice theories we have simulated
are listed in table~1. In the first two runs, $D_6$ and $E_8$, 
only the light quarks, with hopping parameter $\kappa_u$,
were included, while the other two runs
are simulations of 2+1 flavour QCD with strange-quark 
hopping parameter $\kappa_s$.
The quoted values of the O($a$) improvement coefficient $\csw$ 
were determined non-perturbatively in refs.~[\ref{JansenSommer}] and 
[\ref{AokiEtAlIII}], respectively, and the boundary
improvement coefficients $\cG$ and $\cF$ were set to unity.

The basic physical parameters of the simulated lattices
can be inferred from simulation results obtained in 
refs.~[\ref{AokiEtAlI},\ref{AokiEtAlII},\ref{AlphaScale}]
at nearby points in parameter space (see table~2).
We quote these figures without errors and solely
with the intention of giving a rough impression of the physical
situation on the lattices we have considered. 
Table~2 shows that
all lattices are at the edge of the large volume regime of QCD. 
From the point of view of the simulation stability, such lattices
are particularly challenging and we therefore expect 
that stability can more easily be achieved when
one proceeds to simulating larger and finer lattices
(cf.~sect.~1).

\subsection 5.2 Algorithm parameters

In all runs reported here, even-odd preconditioning,
the second kind of twisted-mass determinant
reweighting and a twisted-mass determinant factorization
with three or four masses $\mu_k$ were used (see table~3).
The molecular-dynamics trajectory length $\tau$ was chosen to be
about $2$ on the finer lattices and $1$ on the coarser ones.
Experience suggests 
that shorter trajectory lengths would have a negative
impact on the autocorrelation times of physical quantities
[\ref{SchaeferTauQII}].
The particular values of $\tau$ in table~3 were chosen so as
to get high acceptance rates $\Pacc$ for the fields produced
by the molecular-dynamics evolution.
In the last column of table~3, the numbers $\Ntr$ of trajectories
generated after thermalization are listed.

\topinsert
%Blanke Zahl
\newdimen\digitwidth
\setbox0=\hbox{\rm 0}
\digitwidth=\wd0
\catcode`@=\active
\def@{\kern\digitwidth}
\tablecaption{Simulation parameters} 
\vskip-3.0ex
$$\vbox{\settabs\+&%
                  xxxxx&x&%                         Run
                  xxxx&x&%                          tau
                  xxxxxxxxxxxxxxxxxxxxx&x&%         mu0,..,mun
                  xxxxxxxxxxxxxxxxxx&x&%            Top level integrator
                  xxxxxxx&xx&%                      Pacc
                  xxxxxxx&\cr%                      Ntr
\thicktablerule
\vskip1.2ex
                \+& \hfill Run\hfill
                 && \hfill $\tau$\hfill
                 && \hfill $\mu_0,\ldots,\mu_n$\hfill
                 && \hfill Top level integrator$^*$\hfill
                 && \hfill $\Pacc$\hfill
                 && \hfill $\Ntr$\hfill
                 &\cr
\vskip1.0ex
\thintablerule
\vskip1.2ex
  \+& \hfill $D_6$\hfill
  &&  \hfill $2.0$\hfill
  &&  $\hskip1.9em0.0045,0.01,0.1,1.0$\hfill
  &&  \hfill $\{\hbox{\rm LPFR}\}^{10}$\hfill
  &&  \hfill $0.94$\hfill
  &&  \hfill $1800$\hfill
  &\cr
\vskip0.3ex
  \+& \hfill $E_8$\hfill
  &&  \hfill $1.8$\hfill
  &&  $\hskip1.9em0.0015,0.01,0.1,1.0$\hfill
  &&  \hfill $\{\hbox{\rm OMF}_2\}^{5}$\hfill
  &&  \hfill $0.93$\hfill
  &&  \hfill $@896$\hfill
  &\cr
\vskip0.3ex
  \+& \hfill $I_1$\hfill
  &&  \hfill $1.2$\hfill
  &&  $\hskip1.9em0.0020,0.05,0.5$\hfill
  &&  \hfill $\{\hbox{\rm OMF}_2\}^{4}$\hfill
  &&  \hfill $0.88$\hfill
  &&  \hfill $1224$\hfill
  &\cr
\vskip0.3ex
  \+& \hfill $I_2$\hfill
  &&  \hfill $1.1$\hfill
  &&  $\hskip1.9em0.0012,0.05,0.5$\hfill
  &&  \hfill $\{\hbox{\rm OMF}_2\}^{6}$\hfill
  &&  \hfill $0.90$\hfill
  &&  \hfill $@400$\hfill
  &\cr
\vskip1.2ex
\thicktablerule
\vskip1.0ex
\+&${\vphantom{k}}^*${\footnotefont The parameter $\lambda$ of the ${\rm OMF}_2$
integrator was set to 1/6 [\ref{Omelyan}].}\cr
}
$$
\vskip-2.0ex
\endinsert

The choice of the rational approximation of the strange-quark
determinant in the $2+1$ flavour runs required some experimenting
since the spectral range of $(\Dmshdag\Dmsh)^{1/2}$ is not known a 
priori. Eventually we settled on a Zolotarev rational function 
with $9$ poles, a spectral approximation range $[0.03,6.1]$
and the factorization
\equation{
  \det R^{-1}=\hbox{constant}\times\det\{R_{0,0}^{-1}\}
  \det\{R_{1,1}^{-1}\}\det\{R_{2,2}^{-1}\}\det\{R_{3,8}^{-1}\}
  \enum
}
of the approximate strange-quark determinant (cf.~subsect.~3.2).
The approximation 
error is sufficiently small in this case
to suppress the fluctuations of the reweighting factor 
$\hat{W}_s$ to a level of a few percent.

For the integration of the molecular-dynamics equations, 
a hierarchical integrator with 3 levels was used in all cases,
$\{{\rm OMF}_4\}^1$ being the integrator on the first as well as
on the second level. The top level integrators are listed in 
table~3. Only the force deriving from the gauge action is
integrated at the lowest level and only the smallest forces
(the ones deriving from $\Spft{0}$, 
$\Spf{0,0}$ and $\Spf{1,1}$) at the top level.
Most forces are thus integrated at the first level.

For the solution of the lattice Dirac equation,
the locally deflated SAP preconditioned GCR algorithm [\ref{DFLI},\ref{SAP}]
was employed except at the largest twisted masses,
where we used the CG and
multi-shift CG [\ref{CGM}] algorithms.
The relative residues of the calculated solutions were required to be
less than $10^{-10}$ in the force computations and at most $10^{-11}$
in the computation of the pseudo-fermion actions. With these tolerances,
the reversibility of the numerical integration of the molecular-dynamics 
equations is guaranteed to a precision of about $10^{-9}$ in the
link variables.

\subsection 5.3 Integration instabilities

The molecular-dynamics evolution is probably chaotic and 
is in any case well known to be sensitive to integration inaccuracies.
A manifestation of integration instabilities are large
energy deficits $\Delta H$ at the end of the molecular-dynamics trajectories.
These can be caused
by accidental near-zero modes of the light-quark Dirac operator,
but there exist further sources of instability as well
(loose solver tolerances, for example, or coherent
effects of the higher modes).

Twisted-mass determinant reweighting eliminates the first source
of instability and is therefore expected to have a positive effect
on the stability of the simulations. The energy deficits observed
in our test runs are indeed well behaved (see fig.~1).
In all cases,
values of $|\Delta H|$ significantly larger than 2 are rare
and occur with an estimated probability of at most a few permille.
While such a high level of stability would be difficult to achieve
without low-mode regularization of the quark determinant,
the log-scale frequency splitting of the determinant
no doubt has a stabilizing effect as well and perhaps
also our choice of the molecular-dynamics integrator.

\topinsert
\vbox{
\vskip0.0cm
\centerline{\epsfxsize=13.0 true cm\epsfbox{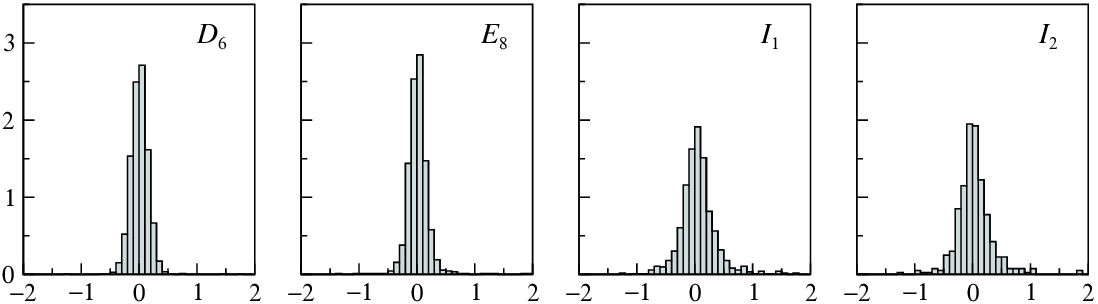}}
\vskip0.3cm
\figurecaption{%
Normalized distribution of the energy deficit $\Delta H$ at the 
end of the molecular-dynamics trajectories as measured in the runs
$D_6,\ldots,I_2$. 
}
%\vskip0.3cm
}
\endinsert

\subsection 5.4 Reweighting efficiency

Stochastic estimates of the light- and strange-quark reweighting factors 
$\hat{W}_2$ and $\hat{W}_s$ can be obtained following the lines of 
subsect.~2.4. In all runs $D_6,\ldots,I_2$ we used $48$ random fields for
the estimation of $\hat{W}_2$ and a single field in the case
of the strange-quark reweighting factor. The latter is actually
nearly constant and little would be gained 
by calculating it more accurately.

\topinsert
\vbox{
\vskip0.0cm
\centerline{\epsfxsize=13.0 true cm\epsfbox{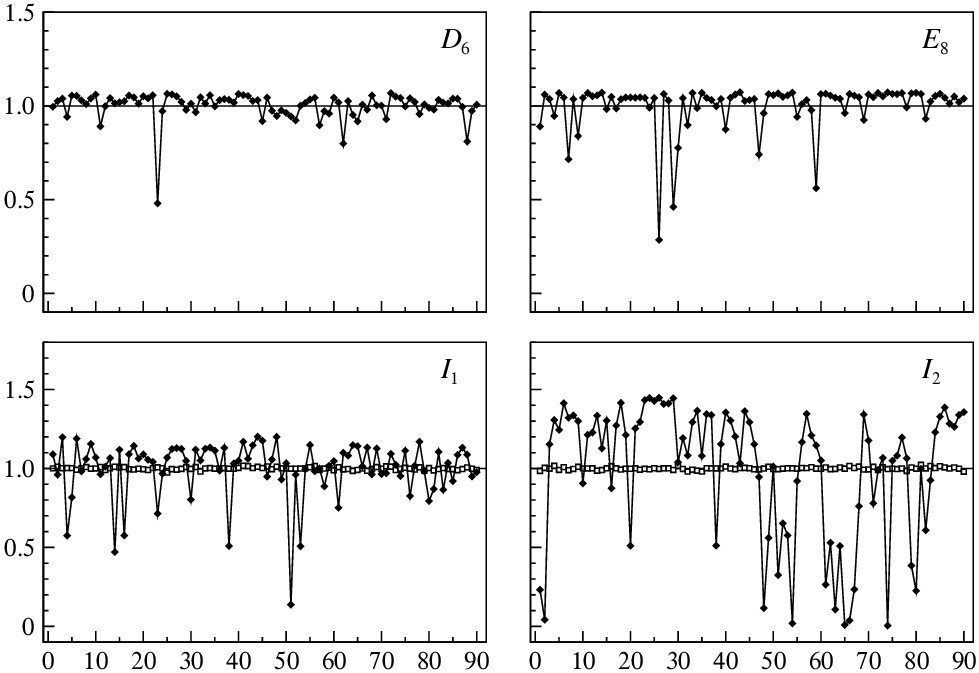}}
\vskip0.3cm
\figurecaption{%
Stochastic estimates of the reweighting factors
$\hat{W}_2$ and $\hat{W}_2\hat{W}_s$ 
(diamonds; connecting lines are drawn to guide the eye)
in the runs $D_6,E_8$ and $I_1,I_2$, respectively, plotted as a function
of the configuration number. 
Configurations are separated by $8$ trajectories
except in run $I_2$, where they are separated by
$4$ trajectories.
In the case of the 2+1 flavour runs, 
the strange-quark reweighting factor 
$\hat{W}_s$ is shown too (squares). 
All reweighting factors plotted in this figure 
are normalized such that their average is equal to $1$.
}
\vskip0.0cm
}
\endinsert

For the reweighting to work out, the normalized 
reweighting factor should remain smaller than $2$ or so, 
as otherwise the ensemble of fields generated in
the simulation is effectively reduced
to the subset of configurations with the dominant weights.
This condition is easily met in all simulations reported here
(see fig.~2). Even in the most critical case, run $I_2$, 
the reweighting factor stays below $1.5$ and 
only $15\%$ of the gauge-field configurations have weight less than $0.5$.
On the simulated lattices and with the chosen values
of the regulator mass $\mu_0$, the efficiency of the simulations 
is thus not compromised by the determinant reweighting.

\subsection 5.5 Low mode sampling

From the point of view of the sampling efficiency,
observables that are sensitive to the low modes
of the light-quark Dirac operator are a special case, because
the series of measured values of such quantities tend 
to have large ``spikes'' at the points in simulation time where the Dirac
operator happens to have near-zero modes. If the data series is 
dominated by a few spikes, a reliable estimation of the 
expectation value of the observable and the associated statistical error
is then practically excluded.

\topinsert
\vbox{
\vskip0.0cm
\centerline{\epsfysize=8.0cm\epsfbox{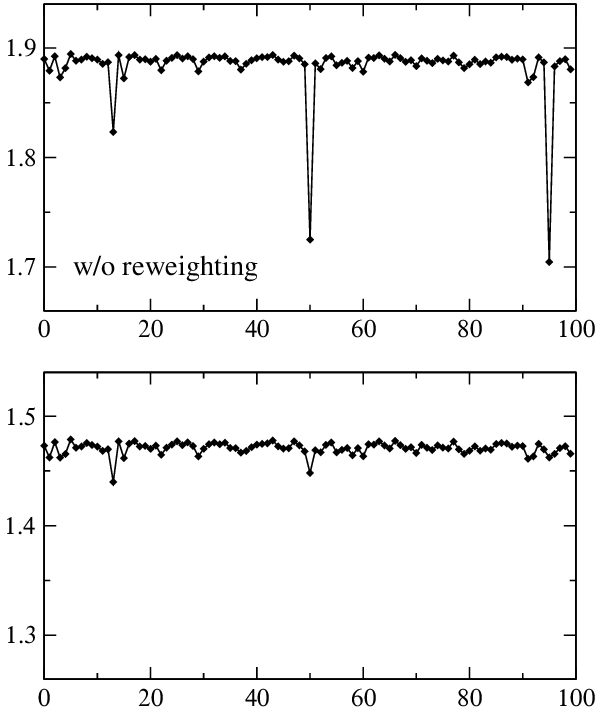}\hskip0.2cm%
            \epsfysize=8.0cm\epsfbox{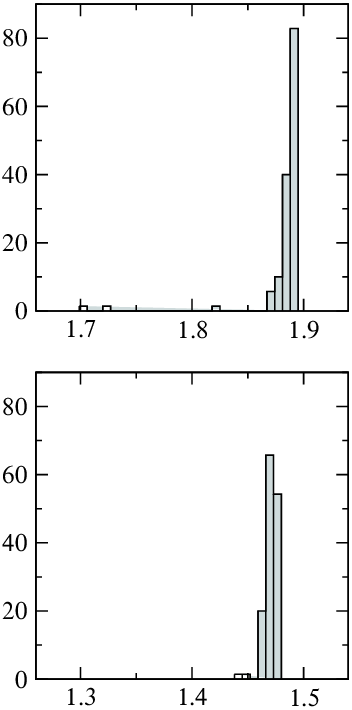}}
\vskip0.3cm
\figurecaption{Jackknife samples of the pion propagator
$G_{\pi}(25,1)$, in lattice units and scaled by $10^3$, 
as measured on the first $100$ gauge-field 
configurations generated in run $I_1$
(configurations are separated by $8$ trajectories).
On the left
the jackknife samples calculated with and without reweighting are plotted
versus the omitted-configuration number.
The normalized distributions of the 
samples are shown on the right.
}
%\vskip0.3cm
}
\endinsert

Exceptionally low eigenvalues
of the Dirac operator can occur
in simulations with twisted-mass reweighting too, but the spikes
in the measurement data series now get suppressed by the 
reweighting factor. The product of the reweighting factor and the
sum of the quark-line diagrams representing
a hadronic correlation function in fact
remains bounded when the Dirac operator becomes singular.

For illustration, consider the pion propagator 
\equation{
  G_{\pi}(x_0,y_0)=-\sum_{\cvec{x}}\langle 
  (\bar{u}\dirac{5}d)(x)(\bar{d}\dirac{5}u)(y)\rangle
  \enum
}
and its computation in run $I_1$ 
at $x_0=25$ and $y_0=1$ (see fig.~3).
In order to reduce the statistical fluctuations, the propagator
was evaluated using $10$ random source fields at time $y_0$.
As can be inferred from the series of the
jackknife samples plotted in the upper left diagram in fig.~3,
the series of measured values has a
few spikes in this run, which are up to $10$ times larger than
the median of the data.

Once the data are reweighted, 
the spikes however disappear, as theoretically anticipated,
and the distribution of the jackknife samples is then entirely 
well behaved
(lower row of diagrams in fig.~3).
Note that the normalized reweighting factor assumes values 
much smaller than $1$ more often than
there are spikes in the propagator data series,
because the propagator is sensitive to only those
low modes of the Dirac operator whose eigenfunctions
have significant support at time $x_0$ and $y_0$.

\subsection 5.6 Simulation cost

Twisted-mass determinant reweighting adds very little to the total 
computational effort required for the simulations. Moreover, whether
the first or the second form of twisted-mass reweighting is used
makes nearly no difference, because the 
additional force term that must be included in the molecular dynamics
evolution in the second case is tiny and can therefore be integrated 
with a large step size.

While the computer time required for the simulations depends
on many parameters, the execution times measured in the runs
reported in this paper may be of some interest.
Using the {\tt openQCD} program [\ref{OQCD}]
and $12$ nodes ($96$ cores) of a standard PC cluster\kern1pt%
\footnote{$\dagger$}{\footnotefont%
The machine used for the tests has
$84$ dual processor nodes with AMD Opteron 2352
($2.1$ GHz quad-core) processors, $8$ GB of DDR2-667 memory
and DDR Infiniband interconnects.},
the computer time required per unit of molecular-dynamics time
in the $D_6$ run was $0.26$ hours.
For the runs $E_8$, $I_1$ and $I_2$, 
we used $32$ nodes (256 cores) of the same machine, the execution
times in these cases being $0.61$, $0.93$ and $1.74$ hours per unit of
molecular-dynamics time.

The true cost of a simulation however also depends on the
autocorrelation times.
As discussed in refs.~[\ref{openQCD},\ref{Villasimius}], 
realistic lower bounds on the exponential
autocorrelation times can be obtained by considering
observables constructed using the Wilson flow [\ref{WilsonFlow}].
The autocorrelation times estimated
in this way turned out to be 
about $32$, $20$ and $15$ in units of molecular-dynamics time
in the $D_6$, $E_8$ and $I_1$ runs, respectively. 
These values are almost a factor $2$ smaller than those determined
at similar lattice spacings
in the SU(3) gauge theory with open boundary conditions
[\ref{openQCD}], but one should keep in mind that the estimates
quoted here are based on much shorter data series and 
may need to be corrected once longer runs are performed.

\section 6. Computation of physical quantities

When open boundary conditions are imposed,
the QCD Hamiltonian and the space of physical states remain
unchanged, but the presence of the boundaries at time $0$ and $T$ 
potentially complicates
the analysis of the calculated correlation functions.
Two cases of interest illustrating the issue
are discussed below.
We focus on run $I_1$ in this section, since 
the physical situation on
this lattice is the one nearest to being
representative of the large-volume regime of QCD.

\subsection 6.1 Reference flow time

Extrapolations to the continuum limit
require the physics on several lattices
to be accurately matched.
The matching can be based on
a comparison of pseudo-scalar meson masses and decay constants, for example,
but as explained in ref.~[\ref{WilsonFlow}],
the finiteness of the Wilson flow in the continuum limit 
[\ref{WilsonFlow},\ref{RenFlow}] may allow
the lattices to be matched far more easily.

A quantity of interest in this context is the reference
flow time $t_0$ implicitly defined through [\ref{WilsonFlow}]
\equation{
  \left\{t^2\langle E(x)\rangle\right\}_{t=t_0}=0.3,
  \qquad E(x)=\frac{1}{4}G^a_{\mu\nu}(x)G^a_{\mu\nu}(x),
  \enum
}
where $G^a_{\mu\nu}(x)$ is a lattice expression for the 
gauge-field tensor at flow time $t$ 
(see appendix B for the definition of the Wilson flow).
On lattices with periodic boundary conditions, 
the expectation value $\langle E(x)\rangle$ is independent of $x$ 
and coincides with its infinite-volume limit up to terms
that vanish exponentially when
lattice sizes $T$ and $L$ are taken to infinity. Note, however, 
that the asymptotic approach to the infinite volume
limit can only be expected to set in at lattice sizes
significantly larger than the smoothing range $\sqrt{8t}$ 
of the Wilson flow.

\topinsert
\vbox{
\vskip0.0cm
\centerline{\epsfxsize=11.0cm\epsfbox{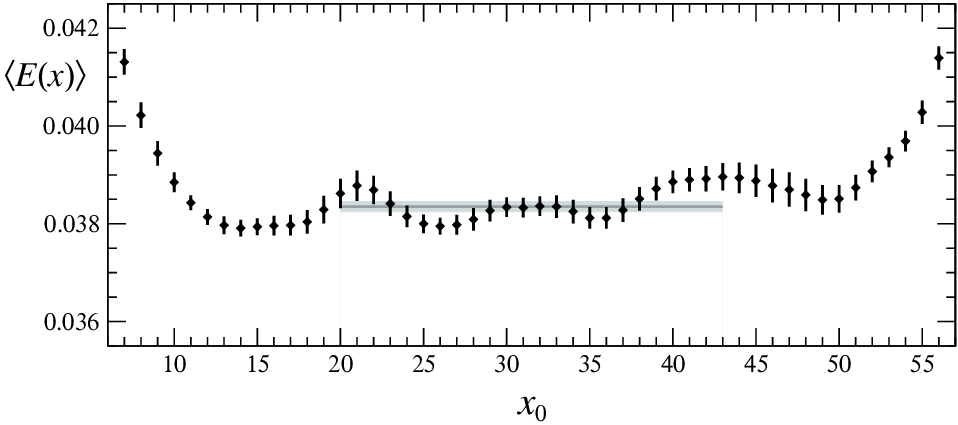}}
\vskip0.3cm
\figurecaption{%
Plot of $\langle E(x)\rangle$ at flow time $t=t_0$,
calculated 
using $150$ gauge-field configurations generated in run $I_1$. 
All quantities are given in lattice units.
Statistical errors were estimated by the jackknife method
after dividing the data into blocks of $6$ consecutive measurements.
The grey line with its error band was obtained
through an uncorrelated least-squares fit
of the data by a constant in the range $20\leq x_0\leq43$.
}
%\vskip0.3cm
}
\endinsert

In the case of open boundary conditions, translation invariance in 
time is broken and $\langle E(x)\rangle$
consequently depends on $x_0$. Similarly to the
finite-volume corrections on periodic lattices,
the effects of the boundaries at time $0$ and $T$
decrease exponentially when one moves away from the boundaries.
On large lattices and at flow times where the smoothing range 
of the Wilson flow is much smaller than the lattice sizes,
$\langle E(x)\rangle$ is thus expected to be
practically constant and equal to its infinite-volume value
in a central range of $x_0$.

\topinsert
\vbox{
\vskip0.0cm
\centerline{\epsfxsize=9.0cm\epsfbox{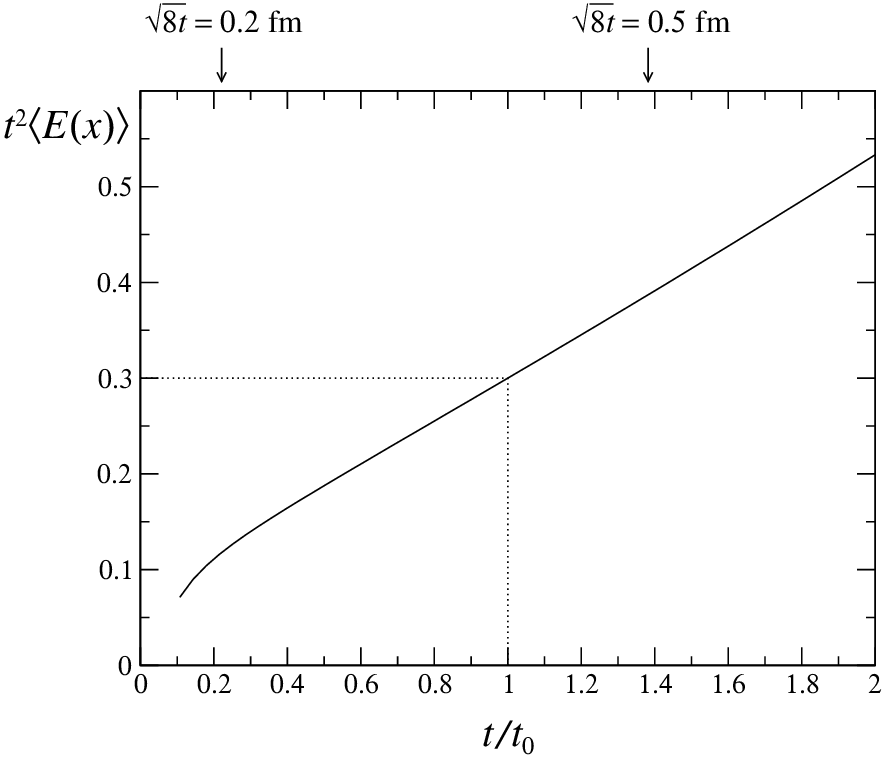}}
\vskip0.3cm
\figurecaption{%
Plot of $t^2\langle E(x)\rangle$
versus the flow time $t$ in units of the reference scale
$t_0$ as measured in the central region of the lattice in run $I_1$.
Statistical errors are too small to be seen on the scale of the plot.
The data at smoothing ranges $\sqrt{8t}$ less than $1.5$ lattice spacings
are strongly affected by lattice effects and are
therefore not shown.
}
%\vskip0.3cm
}
\endinsert

Up to statistical fluctuations of $1-2\%$,
the simulation results for $\langle E(x)\rangle$ 
shown in fig.~4 are indeed
consistent with the existence of a plateau
in a broad range of $x_0$.
In this calculation, a symmetric (clover) expression was
employed for the gauge field tensor $G^a_{\mu\nu}(x)$ 
and the expectation value of $E(x)$
was estimated by averaging
\equation{
   \Ebar(x_0)={1\over4L^3}\sum_{\cvec{x}}
   G^a_{\mu\nu}(x)G^a_{\mu\nu}(x)
   \enum
}
over the gauge fields.
The waves in the central 
region seen in fig.~4 can be explained by recalling
that the Wilson flow suppresses the high-frequency modes of the gauge field. 
The calculated values of $\Ebar(x_0)$ are therefore strongly correlated and 
thus tend to fluctuate coherently over distances
in $x_0$ roughly equal to the smoothing range 
of the flow (which is about $5$ lattice spacings in the case of fig.~4).

From the expectation values $\langle E(x)\rangle$ measured in the 
central region of the lattice, 
the result $t_0=2.792(10)$ is obtained for the reference flow time 
in run $I_1$.
The associated smoothing range, $\sqrt{8t_0}$,
is $4.7$ lattice spacings and $0.43$ fm in physical units.
As a function of the flow time $t$, the behaviour of the dimensionless
combination $t^2\langle E(x)\rangle$ in the range shown in fig.~5
is practically the same as in the pure gauge theory [\ref{WilsonFlow}].
In particular, the combination rises nearly linearly 
beyond a smoothing range of $0.2$ fm or so.

\topinsert
\vbox{
\vskip0.0cm
\centerline{\epsfxsize=9.0cm\epsfbox{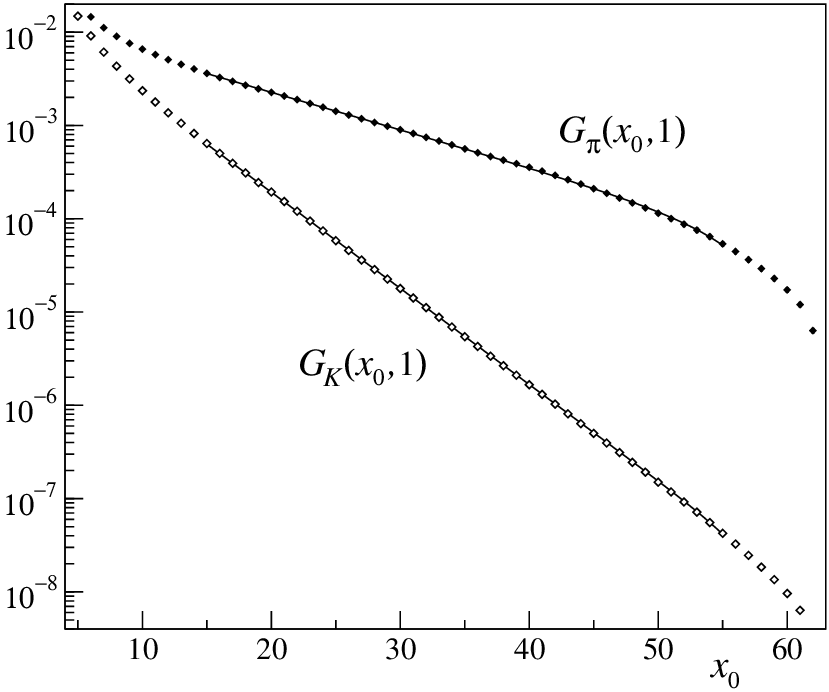}}
\vskip0.3cm
\figurecaption{%
Pion and kaon propagator in lattice units, calculated using an 
ensemble of $150$ gauge-field configurations generated in run $I_1$.
The lines are leading-order chiral perturbation theory fits to the data
(eq.~(6.4) and the corresponding expression for the kaon propagator).
}
%\vskip0.3cm
}
\endinsert

\subsection 6.2 Pseudo-scalar meson masses

The pion mass can be extracted from the pion propagator 
$G_{\pi}(x_0,y_0)$ using standard methods
(see subsect.~5.2 for the definition of the propagator).
In the case shown in fig.~6, the source point is 
at $y_0=1$ and as a function of $x_0$ the propagator 
decreases roughly exponentially from there to the other end of the lattice.
Apart from the fact that it falls off more rapidly,
the kaon propagator behaves essentially in the same way.
In the central region of the lattice,
the effective masses determined from the propagators
are constant within errors and a fit of the data
yields $\mpi=0.0925(19)$ and $\mK=0.2373(10)$ for
the meson masses in lattice units (see fig.~7)\kern1pt%
\footnote{$\dagger$}{\footnotefont%
In physical units, the calculated masses 
($203(4)$ and $520(2)$ MeV, respectively) are slightly smaller
than the values quoted in table~2. 
The probability for the differences to be purely statistical 
is not completely negligible, but
they could also derive from our interpolation of the results 
obtained in refs.~[\ref{AokiEtAlI},\ref{AokiEtAlII}] or from
the presence of finite-volume effects in some of these data.}.

\topinsert
\vbox{
\vskip0.0cm
\centerline{\epsfxsize=9.0cm\epsfbox{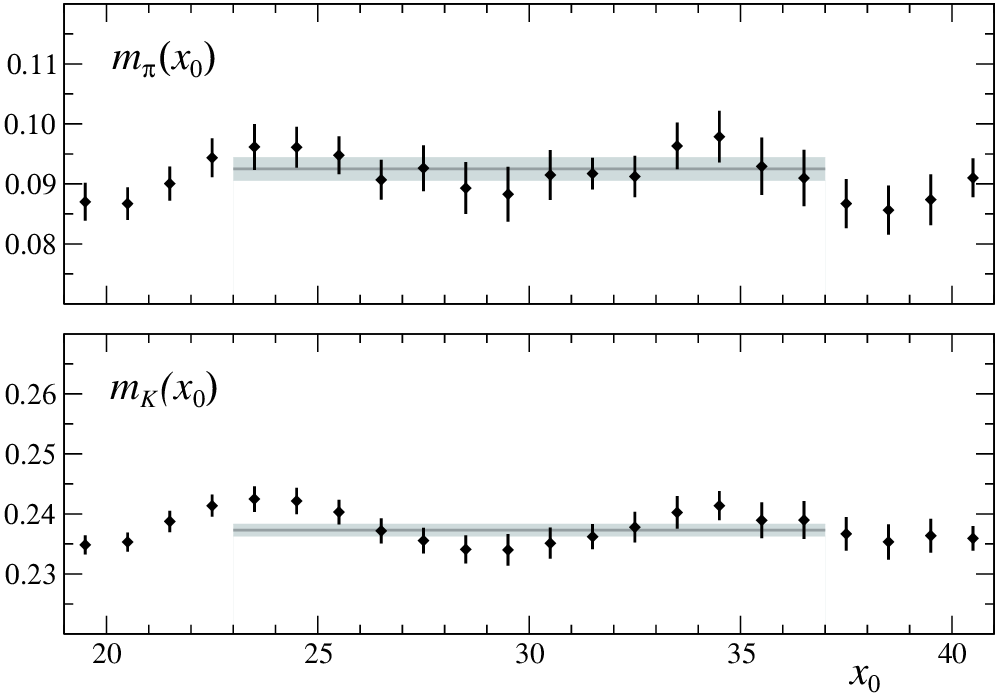}}
\vskip0.3cm
\figurecaption{%
Effective pion and kaon masses measured in the central region
of the lattice in run $I_1$. The grey lines with their error
bands were obtained through uncorrelated 
least-squares fits of the data.
}
%\vskip0.3cm
}
\endinsert

At small times $x_0$, the pseudo-scalar
meson propagators plotted in fig.~6 receive contributions from 
higher-energy intermediate
states as is the case on lattices with periodic boundary 
conditions. Deviations from a single-exponential curve are, however,
also seen when $x_0$ approaches the boundary of the lattice at time $T$.
Close to the chiral limit, and at distances from the boundary 
not smaller than $0.5$ fm or so, these effects will be dominated
by intermediate pseudo-scalar meson states and may therefore conceivably
be described by chiral perturbation theory.

The boundary conditions to be used in chiral perturbation theory
cannot be easily inferred from QCD. Dirichlet
boundary conditions are however known to be
the generic boundary conditions (i.e.~those that do not require
a fine-tuning or a particular symmetry pattern)
in scalar field theories [\ref{Symanzik},\ref{SFbcd}].
Since the flavour symmetry is preserved,
the correct boundary conditions on the pion field
$\pi^a$ (where $a=1,2,3$ is the isospin index) are thus
likely to be
\equation{
   \left.\pi^a(x)\right|_{x_0=0}=\left.\pi^a(x)\right|_{x_0=T}=0.
   \enum
}
Note that these break the axial symmetries, as do the boundary
conditions in the fundamental theory.

To leading order of chiral perturbation theory,
the pion field satisfies the field equation
$(-\Delta+\mpi^2)\pi^a(x)=0$. On-shell correlation functions
of its zero-momentum component
are therefore linear combinations of the exponentials $\exp\{\pm\mpi x_0\}$.
Taking the boundary conditions into account, the pion
propagator plotted in fig.~6 is thus expected to be of the form
\equation{
   G_{\pi}(x_0,1)\propto\sinh(\mpi(T-x_0))
   \enum
}
in the central part of the lattice and at times $x_0$
close (but not too close) to $T$.
Up to higher-order corrections, the same formula, 
with $\mpi$ replaced by $\mK$, should apply in the
case of the kaon propagator.

The leading-order formulae actually fit the propagators quite well
(curves in fig.~6). For the meson masses,
the values quoted above were inserted in these fits and
$T$ was slightly adjusted from $63$ to about $59$
lattice spacings (in the chiral theory,
$T$ is an effective parameter whose value is dynamically determined
by the properties of QCD near the boundaries).
While further confirmation is clearly needed, the 
form of the meson propagators measured in run $I_1$ thus 
lends support to the conjecture that the chiral limit of 
QCD with open boundary conditions is described by the standard
chiral effective theory with Dirichlet boundary conditions.

\section 7. Concluding remarks

The use of open boundary conditions and 
twisted-mass determinant reweighting in numerical
lattice QCD is profitable from the point of view 
stability, efficiency and conceptual clarity.
While open boundary conditions slightly complicate the 
physics analysis of the calculated correlation functions, 
there are currently no practical alternative ways to avoid
the well-known ergodicity problems related to the
emergence of the topological charge sectors 
in the continuum limit.
Twisted-mass determinant reweighting, on the other hand, 
ensures the absence of instabilities and sampling 
inefficiencies caused by accidental near-zero modes
of the lattice Dirac operator
when the Wilson formulation of the 
lattice theory is employed.

The simulation algorithm used in the runs reported 
in this paper combines twisted-mass 
determinant reweighting with
a particular (``log-scale'') Hasenbusch factorization 
[\ref{Hasenbusch},\ref{HasenbuschJansen}]
of the quark determinant
and a hierarchical integrator for the molecular-dynamics
equations based on some of the highly efficient integration
rules proposed by Omelyan, Mryglod and Folk [\ref{Omelyan}].
In all runs and with very little parameter tuning,
an excellent stability and performance of the 
simulations could be achieved in this way.

There is every reason to expect that 
the situation will be essentially unchanged in this
respect when larger and finer lattices
than those considered here are simulated.
The theoretical discussion in ref.~[\ref{TMRW}] moreover
suggests that the reweighting efficiency will depend only
weakly on the lattice parameters if the second kind of 
twisted-mass determinant reweighting is used
with an appropriate choice of the regulator mass.

\vskip1.0ex minus 0.5ex
Most simulations reported in this paper were performed on a
dedicated PC cluster at CERN. We are grateful
to the CERN management for funding this machine
and to the CERN IT Department for technical support.
Thanks also go to the John von Neumann Institute for Computing
for computer time on a Blue Gene/P machine.

\vskip-1.0ex plus 1.0ex

\appendix A. Gauge action

Let $\lps{0}$ and $\lps{1}$ be the sets of oriented
$1\times1$ plaquette and $1\times2$ rectangular loops on the lattice
(the time coordinate $x_0$ of the corners of all these loops 
must thus be in the range $0\leq x_0\leq T$).
The gauge actions considered in this paper are of the form
\equation{
  \SG={1\over g_0^2}\sum_{k=0}^1c_k\sum_{\lp\in\lps{k}}
  w_k(\lp)\,\tr\{1-U(\lp)\},
  \enum
}
where $U(\lp)$ denotes the ordered product of the link variables
$U(x,\mu)$ around $\lp$ and $w_k(\lp)$ is a weight factor
specified below. In order to ensure the correct normalization
of the bare coupling $g_0$, the coefficients $c_k$ must be
such that
\equation{
  c_0+8c_1=1.
  \enum
}
The Wilson plaquette, the tree-level Symanzik-improved [\ref{SymImpI}] 
and the Iwasaki action [\ref{Iwasaki}] are obtained by setting
$c_1=0$, $c_1=-1/12$ and $c_1=-0.331$, respectively. In all cases
the standard convention $\beta=6/g_0^2$ is used for the inverse coupling.

The weight factors $w_k(\lp)$ in eq.~(A.1) are equal to $1$ except
for the space-like loops $\lp$ on the boundaries of the lattice at time $0$
and $T$, where
\equation{
  w_k(\lp)=\frac{1}{2}\cG.
  \enum
}
As previously discussed in ref.~[\ref{openQCD}], the coefficient
$\cG$ is required for $\rmO(a)$ improvement of correlation functions
involving fields close to or at the boundaries of the lattice.
In particular, setting $\cG=1$ ensures on-shell improvement
at tree-level of perturbation theory.

\appendix B. Wilson flow

On lattices with periodic boundary conditions,
the Wilson flow
$V_t(x,\mu)$ of lattice gauge fields is defined by the equations
\equation{
  \partial_t{V}_t(x,\mu)=
  -g_0^2\left\{\partial^a_{x,\mu}\Sw(V_t)\right\}T^aV_t(x,\mu),
  \qquad
  \left.V_t(x,\mu)\right|_{t=0}=U(x,\mu),
  \enum
}
where $\Sw$ denotes the Wilson plaquette action and the parameter 
$t\geq0$ is referred
to as the flow time (see ref.~[\ref{WilsonFlow}] for an introduction
to the subject). 

When open boundary conditions are imposed, 
the flow equation assumes a slightly different form,
\equation{
  \partial_t{V}_t(x,0)=
  -g_0^2\bigl\{\partial^a_{x,0}\SG(V_t)\bigr\}T^aV_t(x,0),
  \quad
  0\leq x_0<T,
  \enum
  \nexteq{2.5ex}
  \partial_t{V}_t(x,k)=
  -{g_0^2\over w(x_0)}
  \bigl\{\partial^a_{x,k}\SG(V_t)\bigr\}T^aV_t(x,k),
  \quad
  0\leq x_0\leq T,
  \quad k=1,2,3,
  \enum
}
where $\SG$ denotes the Wilson action (A.1) with $\cG=1$.
The weight factor
\equation{
  w(x_0)=\cases{\frac{1}{2} & if $x_0=0$ or $x_0=T$,\cr
                 \noalign{\vskip1ex}
                 1           & otherwise,\cr}
  \enum
}
is required to ensure the absence of O($a$) lattice effects in
expectation values of gauge invariant quantities at flow time $t>0$. 

Note that O($a$) improvement is guaranteed if the flow is improved
at tree-level of perturbation theory [\ref{RenFlow}], a property
which can be easily established by calculating the time-dependent 
propagator of the gauge field. The flow equation (B.2)--(B.4) can 
also be derived using an orbifold construction 
previously employed by Taniguchi
[\ref{Taniguchi}] in a different context.
O($a$) lattice effects are then seen to be excluded 
by symmetry.

\beginbibliography

% Open boundary conditions

\bibitem{openQCD}
M. L\"uscher, S. Schaefer,
{\it Lattice QCD without topology barriers},
JHEP 1107 (2011) 036

% Twisted-mass determinant reweighting

\bibitem{TMRW}
M. L\"uscher, F. Palombi,
{\it Fluctuations and reweighting of the quark determinant on 
large lattices}, PoS (LATTICE 2008) 049

% Critical slowing down [SU(3) gauge theory, link update]

\bibitem{DelDebbioTauQ}
L. Del Debbio, H. Panagopoulos, E. Vicari,
{\it $\theta$-dependence of SU(N) gauge theories},
JHEP 08 (2002) 044

% Critical slowing down [HMC algorithm]

\bibitem{SchaeferTauQI}
S. Schaefer, R. Sommer, F. Virotta,
{\it Investigating the critical slowing down of QCD simulations},
PoS (LAT2009) 032

\bibitem{SchaeferTauQII}
S. Schaefer, R. Sommer, F. Virotta,
{\it Critical slowing down and error analysis in lattice QCD simulations},
Nucl. Phys. B845 (2011) 93

% Wilson flow & emergence of the topological sectors

\bibitem{WilsonFlow}
M. L\"uscher,
{\it Properties and uses of the Wilson flow in lattice QCD},
JHEP 1008 (2010) 071

% Topology freezing (review)

\bibitem{Villasimius}
M. L\"uscher,
{\it Topology, the Wilson flow and the HMC algorithm},
PoS (Lattice 2010) 015

% Wilson formulation of lattice QCD

\bibitem{Wilson}
K. G. Wilson, {\it Confinement of quarks}, Phys. Rev. D10 (1974) 2445

% O(a) improvement

\bibitem{SW}
B. Sheikholeslami, R. Wohlert, 
{\it Improved continuum limit lattice action for QCD with Wilson fermions},
Nucl. Phys. B259 (1985) 572

\bibitem{SFimp}
M. L\"uscher, S. Sint, R. Sommer, P. Weisz,
{\it Chiral symmetry and O(a) improvement in lattice QCD},
Nucl. Phys. B478 (1996) 365

% HMC algorithm

\bibitem{HMC}
S. Duane, A. D. Kennedy, B. J. Pendleton, D. Roweth,
{\it Hybrid Monte Carlo},
Phys. Lett. B195 (1987) 216.

% Stability papers

\bibitem{Stability}
L. Del Debbio, L. Giusti, M. L\"uscher, R. Petronzio, N. Tantalo,
{\it Stability of lattice QCD simulations and the thermodynamic limit},
JHEP 0602 (2006) 011

\bibitem{StabilityII}
L. Del Debbio, L. Giusti, M. L\"uscher, R. Petronzio, N. Tantalo,
{\it QCD with light Wilson quarks on fine lattices (II):
DD-HMC simulations and data analysis},
JHEP 0702 (2007) 082

% Eigenvalue distribution in RMT

\bibitem{RMTI}
P. H. Damgaard, K. Splittorff, J. J. M. Verbaarschot,
{\it Microscopic spectrum of the Wilson Dirac operator},
Phys. Rev. Lett. 105 (2010) 162002

\bibitem{RMTII}
G. Akemann, P. H. Damgaard, K. Splittorff, J. J. M. Verbaarschot,
{\it Spectrum of the Wilson Dirac operator at finite lattice spacings},
Phys. Rev. D83 (2011) 085014

\bibitem{RMTIII}
K. Splittorff, J. J. M. Verbaarschot,
{\it The Wilson Dirac spectrum for QCD with dynamical quarks},
Phys. Rev. D84 (2011) 065031

\bibitem{RMTIV}
M. Kieburg, K. Splittorff, J. J. M. Verbaarschot,
{\it The realization of the Sharpe-Singleton scenario},
arXiv:1202.0620v1

% First study of twisted-mass reweighting

\bibitem{MiaoEtAl}
C. Miao, H. B. Meyer, H. Wittig,
{\it Twisted-mass reweighting for O(a) improved Wilson fermions},
PoS (Lattice 2011) 041

% PACS-CS physical point simulations

\bibitem{AokiEtAlI}
S. Aoki et al.~(PACS-CS collab.), 
{\it 2+1 flavor lattice QCD toward the physical point},
Phys. Rev. D79 (2009) 034503

\bibitem{AokiEtAlII}
S. Aoki et al.~(PACS-CS collab.),
{\it Physical point simulation in 2+1 flavor lattice QCD},
Phys. Rev. D81 (2010) 074503

% openQCD program

\bibitem{OQCD}
{\tt http://cern.ch/luscher/openQCD}

% Computation of reweighting factors

\bibitem{MRW}
A. Hasenfratz, R. Hoffmann, S. Schaefer,
{\it Reweighting towards the chiral limit},
Phys. Rev. D78 (2008) 014515

% Hasenbusch factorization

\bibitem{Hasenbusch}
M. Hasenbusch,
{\it Speeding up the Hybrid Monte Carlo algorithm for dynamical fermions},
Phys. Lett. B519 (2001) 177

\bibitem{HasenbuschJansen}
M. Hasenbusch, K. Jansen,
{\it Speeding up lattice QCD simulations with clover-improved Wilson 
fermions}, Nucl. Phys. B659 (2003) 299

\bibitem{UrbachEtAl}
C. Urbach, K. Jansen, A. Shindler, U. Wenger,
{\it HMC algorithm with multiple time scale 
integration and mass preconditioning},
Comp. Phys. Commun. 174 (2006) 87

% DDHMC

\bibitem{DDHMC}
M. L\"uscher,
{\it Schwarz-preconditioned HMC algorithm for two-flavor lattice QCD},
Comp. Phys. Commun. 165 (2005) 199

% RHMC algorithm

\bibitem{RHMCI}
I. Horvath, A. D. Kennedy, S. Sint,
{\it A new exact method for dynamical fermion computations with
non-local actions},
Nucl. Phys. (Proc. Suppl.) 73 (1999) 834

\bibitem{RHMCII}
M. A. Clark, A. D. Kennedy,
{\it Accelerating dynamical fermion computations using the
Rational Hybrid Monte Carlo (RHMC) algorithm with
multiple pseudo-fermion fields},
Phys. Rev. Lett. 98 (2007) 051601

% Les Houches lectures

\bibitem{LesHouches}
M.~L\"uscher,
{\it Computational strategies in lattice QCD},
in: {\it Modern perspectives in lattice QCD},
eds. L. Lellouch et al. (Oxford University Press, New York, 2011)
[arXiv: 1002.4232]

% Zolotarev rational functions

\bibitem{Achiezer}
N. I. Achiezer,
{\it Theory of approximation}
(Dover Publications, New York, 1992)

% Multiple time scale integration

\bibitem{SextonWeingarten}
J. C. Sexton, D. H. Weingarten,
{\it Hamiltonian evolution for the Hybrid Monte Carlo algorithm},
Nucl. Phys. B380 (1992) 665

% 2nd and 4th order Omelyan integrator

\bibitem{Omelyan}
I. P. Omelyan, I. M. Mryglod, R. Folk,
{\it Symplectic analytically integrable decomposition
algorithms: classification, derivation, and application to molecular
dynamics, quantum and celestial mechanics simulations},
Comp. Phys. Commun. 151 (2003) 272

% Local deflation

\bibitem{DFLI}
M. L\"uscher,
{\it  Local coherence and deflation of the low quark modes in lattice QCD},
JHEP 0707 (2007) 081

\bibitem{DFLII}
M. L\"uscher,
{\it  Deflation acceleration of lattice QCD simulations},
JHEP 0712\hfill\break (2007) 011

% Multi-shift CG

\bibitem{CGM}
B. Jegerlehner,
{\it Krylov space solvers for shifted linear systems},
preprint IUHET-353 (1996) [arXiv: hep-lat/9612014]

% SAP preconditioning

\bibitem{SAP}
M. L\"uscher,
{\it Solution of the Dirac equation in lattice QCD using a
domain decomposition method},
Comp. Phys. Commun. 156 (2004) 209

% Non-perturbative O(a) improvement

\bibitem{JansenSommer}
K. Jansen, R. Sommer,
{\it O(a) improvement of lattice QCD with two flavors of Wilson quarks},
Nucl. Phys. B530 (1998) 185 [E: {\it ibid.} B643 (2002) 517]

\bibitem{AokiEtAlIII}
S. Aoki et al. (CP-PACS collab.),
{\it Nonperturbative O(a) improvement of the Wilson quark action with the 
RG-improved gauge action using the Schr\"odinger functional method},
Phys. Rev. D73 (2006) 034501

% Scale setting in two-flavour QCD

\bibitem{AlphaScale}
P. Fritzsch et al. (ALPHA collab.),
{\it The strange quark mass and Lambda parameter of two flavor QCD},
arXiv:1205.5380

% Renormalization of the gradient flow

\bibitem{RenFlow}
M. L\"uscher, P. Weisz,
{\it Perturbative analysis of the gradient flow in non-Abelian gauge
theories},
JHEP 1102 (2011) 051

% Renormalization of quantum field theories on a half space

\bibitem{Symanzik}
K. Symanzik,
{\it Schr\"odinger representation and Casimir effect in
renormalizable quantum field theory},
Nucl. Phys. B190 [FS3] (1981) 1

\bibitem{SFbcd}
M. L\"uscher,
{\it The Schr\"odinger functional in lattice QCD with exact chiral symmetry},
JHEP05(2006)042

% Symanzik improved action

\bibitem{SymImpI}
P. Weisz,
{\it Continuum limit improved lattice action for pure Yang-Mills theory (I)},
Nucl. Phys. B212 (1983) 1

% Iwasaki action

\bibitem{Iwasaki}
Y. Iwasaki,
{\it Renormalization group analysis of lattice theories and
improved lattice action. II -- Four-dimensional non-Abelian SU(N) gauge model},
preprint UTHEP-118 (1983) and arXiv:1111.7054v1

% Orbifold construction

\bibitem{Taniguchi}
Y. Taniguchi,
{\it Schr\"odinger functional formalism with Ginsparg--Wilson fermion},
JHEP 0512 (2005) 037;

\endbibliography

\bye